# SUGAR-R: Robust Online Restoration Platform for SCADA-Absent Grid

Amritanshu Pandey[1], Member, Aayushya Agrawal[1], Graduate Student Member, Marko Jereminov[1], Member, Larry Pileggi[1], Fellow, *IEEE*

*Abstract*— Secure and fast grid restoration from a collapsed state is increasingly critical as blackouts are becoming more prevalent around the globe. Generally, the restoration of grid during a blackout is achieved with the help of Supervisory Control and Data Acquisition System (SCADA) based central control; however, with the threat of cyber-blackouts, this presumption of an available and secure SCADA system is not valid. This is also true for grids in developing countries as well as for many distribution and micro grids that lack SCADA. In this paper, we introduce an online framework for localized grid restoration that validates and updates a pre-defined crank path in real-time based on the vital grid states of voltages, currents and frequency. The proposed framework maintains an online network topology of the localized grid that can continuously sample local measurements and update the grid model, thereby circumventing SCADA based central control. In the results section we demonstrate the efficacy of this framework for blackstart by ensuring a feasible crank path with voltage and frequency states within bounds, while further assisting in synchronization of two disconnected sub-grids during the re-energization process using a distributed framework.

## Nomenclature

| | |
|---|---|
| $I_{D,BIG}^R, I_{D,BIG}^I$ | Current consumption by a BIG load model. |
| $\alpha_{BIG}^R, \alpha_{BIG}^I$ | Constant current part of the BIG load model. |
| $G_{BIG}, B_{BIG}$ | Linear sensitivities of the BIG load model. |
| $I_{G,i}^R, I_{G,i}^I$ | Real and imaginary currents injections by a generator. |
| $I_{D,i}^R, I_{D,i}^I$ | Real and imaginary currents injections by a load. |
| $V_i^R, V_i^I$ | Real and imaginary voltage at node $i$. |
| $G_{ik}^Y, B_{ik}^Y$ | Components of admittance matrix. |
| $P_g^{C_t}, Q_g^{C_t}$ | Real and reactive power of a generator at crank path step $C_t$. |
| $P_d^D, Q_d^D$ | Real and reactive power of a constant power demand. |
| $\Delta P_g^{C_t}$ | Change in real power set-point of the generator for crank path step $C_t$. |
| $\Delta P_P^{C_t}$ | Change in real power of generator due to droop characteristics for step $C_t$. |
| $\Delta f$ | Delta frequency change in the network. |
| $\Delta V_i^{SQ}$ | Square of deviation from set voltage for PV buses during synchronization. |
| $M$ | Inertia component of a generator. |
| $\overline{\mu}, \underline{\mu}$ | Vector of dual variables corresponding to the upper and lower bound of states vector $X$, respectively. |
| $\lambda$ | Vector of dual variables corresponding to the equality constraint. |
| $X, \overline{X}, \underline{X}$ | System state vector along with the vector of upper and lower bounds, respectively. |

## I. Introduction

Blackouts are becoming prevalent threats around the globe affecting developed centralized grids as well as weak fragmented grids alike. The advent of cyber-attacks [1] have further escalated the risk of blackouts in developed grids, while fragile grids remain susceptible to rolling blackouts due to a lack of energy infrastructure [2]. In most instances, grid restoration following a blackout is driven by SCADA-based central control, however during cyber-attacks the SCADA and Energy Management Systems (EMS) may become inoperable or even worse, compromised [3]. Presently, restoration without a SCADA and EMS requires experienced engineers to manually guide the process. However, such manual operation may result in sub-optimal restoration process or damage to equipment. Additionally, it may lead to delayed energization of critical resources such as hospitals, military installations, and transport systems. Due to these drawbacks, both regulatory bodies and system operators are interested in methodologies that can help in restoration of grids under black start with no SCADA [4]-[6] and they underscore the need for robust simulation frameworks that can help achieve these goals. This was clearly stated in a report by the Federal Energy Regulatory Commission (FERC) in [4].

In search of that goal, this paper develops a novel online localized and distributed modeling and simulation framework that can assist in *black starting* a grid with a compromised or unavailable SCADA network. We term the framework as Simulation with Unified Grid Analyses and Renewables – Restoration (SUGAR-R) and is shown in Fig. 1.

This framework will be deployed in critical distributed locations (e.g. Area 1 and 2 in Fig. 1) in the grid and will maintain an online network topology of the localized network by continuously sampling various classes of local measurements (see sensors in Fig. 1) and mapping them to system models. These measurements will not be provided via the central SCADA system, rather from the localized meters through secondary channels (see [7]), thereby avoiding invalid system models caused by potential false data injection cyber-attacks on the central EMS based system. In the restoration process studied by FERC, collecting state measurements in SCADA-less environment is presently done using personnel

This work was supported in part by the Defense Advanced Research Projects Agency (DARPA) under award no. FA8750-17-1-0059 for the RADICS program.
[1]Authors are with the Electrical and Computer Engineering Department, Carnegie Mellon University, Pittsburgh, PA 15213 USA, (e-mail: {amritanp, aayushya, mjeremin, pileggi}@andrew.cmu.edu).

support [4]; however, with the advent of IOTs and cloud computing, localized and cheap autonomous data collection is possible [7]. Nonetheless, this paper is not poised to determine the acquisition of state measurements in a SCADA absent system, but rather to use available local measurements to assist in grid restoration.

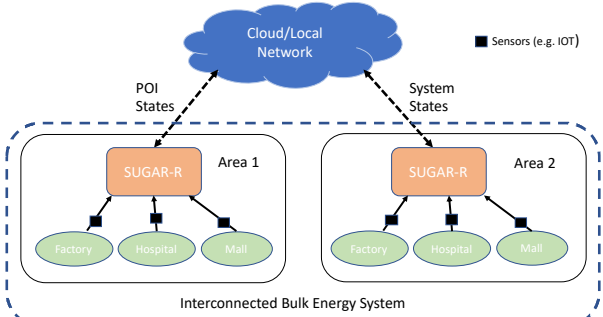

*Figure 1: Pictorial representation of SUGAR-R framework.*

The proposed online framework will assist in restoration by simulating the energization of sections of the localized grid in steady-state stages, known as the crank path. For a successful restoration process, the system during each crank step should maintain *feasibility and stability* by ensuring that vital health states (i.e. voltage, currents, and frequency) of the network are within operational bounds through proper matching of load and generation. SUGAR-R will validate these criteria and recommend optimal corrective actions in case no *feasible* state is possible based on the pre-defined crank path. Once an island is fully restored, SUGAR-R will use a proposed distributed computing methodology with cloud or local interface to assist in synchronization of multiple islanded networks [4].

SUGAR-R utilizes the physics of the grid and provides a lightweight and robust framework for grid simulation that is apt for studying online grid restoration of distributed grid segments. It is able to better model system states by representing physical characteristics of the grid through natural state variables of voltage, current and frequency, unlike other previous works. For instance, [8]-[11] considers loss-less networks or uses DC approximations to represent the network flows (neglecting the voltage characteristics of the line), and often steady-state methodologies [8],[12]-[13] do not consider the impact of frequency. Frequency is an important variable to model while simulating grid failure modes, without which the effect of frequency-based control actions cannot be predicted. For instance, a recent power failure in the U.K. grid due to an N-2 contingency, triggered a five percent drop of the system load by an automatic under-frequency load shedding (UFLS) scheme that actuated at 48.8 Hz [14]. A steady-state simulation framework without frequency as a variable will not able to accurately simulate this event.

To robustly incorporate the effects of frequency, SUGAR-R extends the prior work of SUGAR [15]-[16], a robust power grid steady-state analysis and optimization framework based on circuit-theoretic approach. The circuit formalism within SUGAR allows inclusion of any physical measurement information into the network model in real-time. In doing so, an operation engineer can expect to use SUGAR-R as a tool during the restoration process to validate and recommend modifications to the crank-path. As SUGAR-R is not only scalable but also operational in real-time, the operation engineer can turn to the tool to ensure that the grid will be stable after any restorative action they are about to perform. After restoring their local network, the operation engineer can then use SUGAR-R to propose a change in dispatch to reliably synchronize with other energized networks.

The remainder of the paper is structured as follows; Section II describes the circuit-theoretic components required for modeling and maintaining the online network model of the localized grid sections. Section III describes the optimization-based grid blackstart framework and demonstrates the distributed method that assists in synchronizing disconnected islands. Finally, the simulation results are discussed in Section V.

## II. A Circuit-Based Localized Network Framework

The goal of a simulation framework during grid restoration without central SCADA is to validate the feasibility and stability of each crank path while also assisting in synchronization of different sections of the grid. In order to achieve the following criteria, the proposed simulation framework must:

- introduce frequency as a state within the framework,
- clearly distinguish a *hard-to-solve* case from an *infeasible (past the tip of nose curve)* case independent of the choice of initial conditions,
- incorporate online local measurements into the network model in real-time to provide accurate snapshot of the localized grid,
- under an *infeasible* scenario, recommend optimal change in device settings to ensure that the voltage and frequency remain within bounds while satisfying KCLs at each node.

To meet these criteria, this paper develops an online framework that can run light-weight optimization analyses while including the frequency state. We will assume that the framework can collect and accordingly update localized network models in real-time (for more details see Section II-B). To build this approach, we will begin with and extend a previously developed circuit theoretic grid analysis framework SUGAR. SUGAR approach described in [15]-[16], is already capable of identifying an infeasible grid state (KCL not satisfied at each node while respecting device limits) from an hard-to-solve network [17].

### A. Frequency Steady-State Framework

To robustly model the impact of frequency controls on the steady state achieved at each crank stage, SUGAR approach was further extended to integrate a frequency deviation term, $\Delta f$ into power flow while implicitly modeling active power limits without the use of outer loops during simulation. This enables accurate and robust modeling of frequency-dependent grid components and controls including the primary frequency response of generators based on the change frequency $\Delta f$:

$$\Delta P_P = -\frac{P_R}{R}\Delta f \qquad (1)$$

where, $\Delta P_P$ is the frequency-adjusted change in set real power output of the generator, $P_g^{SET}$, and $P_R$ and $R$ define the primary frequency response of the generator based on droop control

[18]. In this approach, the primary response is also extended for slack generators, thereby constraining the slack power to a realistic output.

In addition, the authors in [19] have implicitly modeled the active power limits of a generator using a continuous function, as shown in Fig. 2. This implicit model is represented by a smooth first-order continuous function (with five regions) that eliminates the use of outer loops to model the generator active power limits, to significantly improve convergence and allow extendibility to optimization framework without the use of complementarity constraints. The proposed SUGAR-R framework will be built upon this work and is further described in following sections.

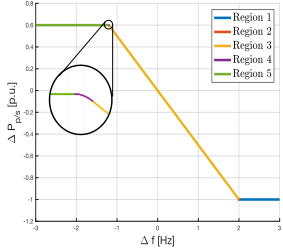

*Figure 2: Implicit model for change in real power of the generator due to primary droop control [19].*

### B. Use of Localized Measurement for Network Modeling

To accurately model the steady state of the power grid, grid operators use SCADA to continuously gather all measurements and device statuses in a central location. With the help of state estimation and topology estimation, grid operators update or develop network models and incorporate them into simulations to further study real-time operations including those for grid restoration. However, under a cyber-attack, central SCADA or EMS systems used for maintaining and updating network model may be corrupted or compromised requiring an alternate methodology for network model creation and updating.

In our approach, we propose the use of localized measurements to alleviate the likelihood of widespread disruption due to one central attack. Any such proposed framework will have to utilize local measurements obtained through secondary channels (independent of central SCADA system), such as smart meters and other devices to maintain and update localized network models. Newer frameworks that are independent of standard SCADA have been successful in capturing grid data cheaply from wide array of measurements and making it available to a common platform [7]. For instance, many such platforms were introduced in a recent federal program focused on restoring the grid under a cyber-attack [20]. Such platforms by collecting grid information of the local network in real-time will provide the grid operators situational awareness during restoration process [21]-[22]. However, the implementation for access to such localized measurements is not the primary goal of this paper, rather it is assumed that the proposed SUGAR-R framework can directly incorporate these measurements into its models.

With access to these measurements, SUGAR's circuit-theoretic framework is well suited to mapping raw measurements into circuit-models using light-weight compute resources with minimum processing and memory requirements (e.g. raspberry pi[24]). These circuit-models used in SUGAR-R allow for physically realistic representations described using natural state variables of voltages, currents and frequency. To demonstrate this, we consider a linear BIG load model [25]-[26] as an example (see Fig. 3).

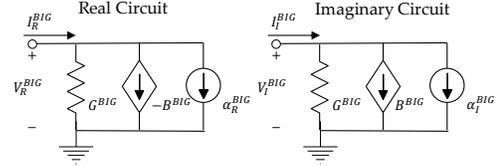

*Figure 3. BIG Load Model.*

Similar to a ZIP and PQ model, but more naturally representing the true sensitivities with respect to voltages across the load, the BIG model is defined by a parallel constant current source ($\alpha_{BIG}^R + j\alpha_{BIG}^I$), conductance ($G_{BIG}$) and susceptance ($B_{BIG}$) and whose real and imaginary currents are given by (2). The conductance and susceptance reflect the linear sensitivities of the load with respect to in-phase and out of phase voltages components.

$$I_{D,BIG}^R + jI_{D,BIG}^I = \alpha_{BIG}^R + j\alpha_{BIG}^I \\ + (V_{BIG}^R + jV_{BIG}^I)(G_{BIG} + jB_{BIG}) \quad (2)$$

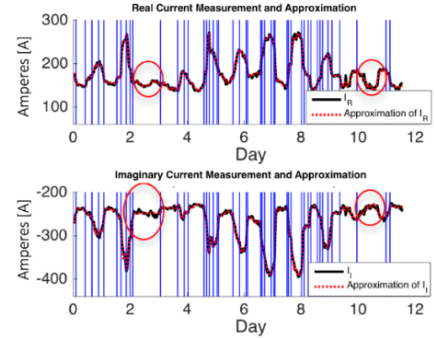

*Figure 4: BIG model representation of 12 days for the LBNL data; BIG model is clearly shown to capture the voltage variability of the load characteristics [26].*

Using techniques from [25], the BIG load model parameters can be fit with online streaming measurement data from a large variety of measurement devices (such as the smart meters, remote terminal units, micro phasor measurement units and other IOTs) to accurately represent the nominal characteristics of the load while also capturing its voltage sensitivity. Moreover, these load models can be aggregated (in parallel) to create larger bulk loads models, without loss of generality, while preserving both nominal as well as the sensitivity behavior of the load.

For instance, Fig. 4 represents the comparison between load currents produced from the fitted load model (shown in red dots using quasi-steady state analysis) and those from true measurements (shown in black line) using data from a sensor at Lawrence Berkeley National Laboratory (LBNL) [27] for a total of 12 days. For each time instance that is indicated by the blue line in Fig. 4, the BIG load model parameters are updated in real-time to capture the variation in load. Not only does the fitted BIG load capture the nominal measured data, but it also models the voltage-sensitive nature of the load currents, as is shown by the red circles in the Fig. 4.

Following the same approach, through online sampling of measurement data, the SUGAR-R approach can also update other elements in the network such as generation and shunts while also updating the branch and switch positions based on the device statuses.

## C. Development of crank path

In [8]-[10], extensive literature is presented on developing the crank path for grid restoration processes. The methods that exist today develop these crank-paths offline. However, during a blackout, equipment may be damaged and network topology may be altered hence alternating the original crank-path and making the resultant off-line analysis inaccurate. Moreover, real-time operation may observe load distributions that are significantly different than the ones that were originally evaluated and will result in a modification of the crank paths. Therefore, even though our proposed methodology starts with a planned crank-path, its following step updates are further based on the real-time decision analysis, available measurement data and network topology all captured within an optimization framework outlined in SUGAR-R.

## III. OPTIMIZATION-BASED SUGAR-R FRAMEWORK

### A. General Methodology

In this section, we describe the internals and methodology of the SUGAR-R optimization framework. The goal of the framework is to validate the feasible actuation of each crank path step while ensuring that the voltage and frequency bounds are enforced, and device limits met. To ensure feasible actuation, the optimization framework minimally adjusts the generator active power set-points during ramp up from what is initially outlined by the pre-defined crank-path. The mathematical formulation of the optimization problem that is solved for the validity of each crank path step ($C_t \in \{C_1, C_2, \ldots, C_n\}$) is defined below:

$$\min_{\Delta P_{g(\forall g \in \mathcal{G})}^{C_t}} \left\| \Delta P_g^{C_t} \right\|_2^2 \tag{3}$$

subject to:

$$I_{G,i}^R - I_{D,i}^R = \mathbf{Re}\left\{ \sum_{k=1}^{|B|} (V_{ik}^R + jV_{ik}^I)(G_{ik}^Y + jB_{ik}^Y) \right\} \quad \forall i \in B \tag{4}$$

$$I_{G,i}^I - I_{D,i}^I = \mathbf{Im}\left\{ \sum_{k=1}^{|B|} (V_{ik}^R + jV_{ik}^I)(G_{ik}^Y + jB_{ik}^Y) \right\} \quad \forall i \in B \tag{5}$$

$$\underline{P_g} \leq P_g^{C_t} + \Delta P_P^{C_t} + \Delta P_g^{C_t} \leq \overline{P_g} \quad \forall g \in \mathcal{G} \tag{6}$$

$$\underline{Q_g} \leq Q_g^{C_t} \leq \overline{Q_g} \quad \forall g \in \mathcal{G} \tag{7}$$

$$\left(\underline{V_i}\right)^2 \leq V_i^{SQ} \leq \left(\overline{V_i}\right)^2 \quad \forall i \in \mathcal{N} \tag{8}$$

$$\underline{\Delta f} \leq \Delta f \leq \overline{\Delta f} \tag{9}$$

$$V_i^{SQ} = V_R^2 + V_I^2 \quad \forall g \in B \tag{10}$$

$$I_{G,i}^R = \sum_{g=1}^{|\mathcal{G}(i)|} \frac{(P_g^{C_t} + \Delta P_g^{C_t} + \Delta P_P^{C_t})V_i^R + Q_g^G V_i^I}{V_{R,i}^2 + V_{I,i}^2} \quad \forall i \in B \tag{11}$$

$$I_{G,i}^I = \sum_{g=1}^{|\mathcal{G}(i)|} \frac{(P_g^{C_t} + \Delta P_g^{C_t} + \Delta P_P^{C_t})V_i^I - Q_g^G V_i^R}{V_{R,i}^2 + V_{I,i}^2} \quad \forall i \in B \tag{12}$$

$$I_{D,i}^R = \sum_{d=1}^{|D(i)|} G_{BIG,d}^{C_t} V_i^R - B_{BIG,d}^{C_t} V_i^I + \alpha_{BIG}^R \quad \forall i \in B \tag{13}$$

$$I_{D,i}^I = \sum_{d=1}^{|D(i)|} G_{BIG,d}^{C_t} V_i^I + B_{BIG,d}^{C_t} V_i^R + \alpha_{BIG}^I \quad \forall i \in B \tag{14}$$

$$\Delta P_P^{C_t} = -M \Delta f \quad \forall g \in \mathcal{G} \tag{15}$$

$$\underline{\Delta P_{ramp}} \leq (P_g^{C_t} + \Delta P_g^{C_t}) - (P_g^{C_{t-1}} + \Delta P_g^{C_{t-1}}) \quad \forall g \in \mathcal{G}$$
$$\leq \overline{\Delta P_{ramp}} \quad \forall C_t \in C \tag{16}$$

The active power $P_g^G$ represents the specified active power from the pre-defined crank-path. Given a set of non-linear equations representing the frequency dependent network equations (4) and (5), an additional active power generation, $\Delta P_g^{C_t}$ at each participating generator ($g \in \mathcal{G}$) is required to ensure feasibility of the network while respecting voltage ($\sqrt{V^{SQ}}$) and frequency ($\Delta f$) bounds given by (8)-(9). The added generation, $\Delta P_g^{C_t}$, represents the delta change of active generation ($g \in \mathcal{G}$) that is required to maintain the health of the system while maintaining device limits (6)-(7). To ensure that the adjustment to each crank-path stage is operational, we also constrain the additional active power to obey ramping constraints given by (16). The ramping constraint ensures that the adjustments at a crank-path stage, $C_t$, are viable with respect to minimum and maximum ramping constraints, $\underline{\Delta P_{ramp}}$ and $\overline{\Delta P_{ramp}}$ respectively.

To solve the defined optimization sequentially for each crank-path stage, $C_t$, we formulate the problem using Primal-Dual Interior Point (PDIP) algorithm [30] and apply a circuit-theoretic approach to solve for the local optima [16]. It is worthwhile to note that a local optimum solution to the optimization problem is sufficient to meet the purposes of this grid restoration approach i.e. to obtain a feasible operating point. To obtain a solution, the circuit-based approach maps the first-order optimality conditions (KKT conditions) for the PDIP problem into a set of equivalent circuits (see [16] for methodology) with following components:

$$\mathcal{L} = \left\| \Delta P_g^{C_t} \right\|_2^2 + \lambda^T g(X) + \mu^T X \tag{17}$$

$$\nabla_\lambda \mathcal{L} = g(X) = 0 \tag{18}$$

$$\nabla_X \mathcal{L} = \nabla_X^T g(X) \lambda = 0 \tag{19}$$

$$\overline{\mu} \odot (X - \overline{X}) + \epsilon = 0 \tag{20}$$

$$-\underline{\mu} \odot (X - \underline{X}) + \epsilon = 0 \tag{21}$$

$$\mu > 0 \tag{22}$$

$$\underline{X} < X < \overline{X} \tag{23}$$

where, $X: \{P_g^G, Q_g^G, V_i^R, V_i^I, \Delta f, \Delta P_G^{C_t}, \Delta P_P^{C_t}\}$ represents the set of system states and $g(X)$ represents the set of equality constraints, and $\underline{X} < X < \overline{X}$ represent the box constraints bounding the system and device variables.

The mapped circuits represent the KKT optimality conditions (18)-(19) along with complementarity slackness constraints corresponding to the inequalities (20)-(21). We solve the non-linear equations corresponding to these circuits using Newton's method [30]. To ensure a feasible convergence, SUGAR-R employs circuit-based homotopy and limiting methods developed in [15] that solve power grid based nonlinear constraints using circuit heuristic methods.

To further improve robustness compared to the existing state-of-art nonlinear optimization algorithms [28]-[29], each variable during Newton-Raphson is separately limited based on the physical characteristics of the problem (see [15], [16]) significantly improving the convergence characteristics. Additionally, to solve highly non-linear equations corresponding to the complimentary slackness conditions (20)-(21), diode limiting that is further described below is implemented and used.

### B. Diode Limiting

To ensure device limits are always met, inequalities of optimization problem given by (6)-(9) are modeled with relaxed complementary slackness constraints [30]. This mathematical representation mimics a diode behavior when constrained to its feasible space (with (25) and (26) satisfied) and is represented by the mathematical form:

$$\mu(x - \overline{x}) + \epsilon = 0 \quad (24)$$

$$\mu > 0 \quad (25)$$

$$x - \overline{x} < 0 \quad (26)$$

where $\mu$ is the slack variable associated to the variable $x$ being bounded by $\overline{x}$. In addition, $\epsilon$ is a small scalar (approximately 1e-6) that ensures a continuous behavior. Following a general methodology in circuit simulation (also see [16]), during Newton Raphson (NR), each iteration will update the dual variable and primal variable by:

$$\mu^{k+1} = \mu^k + \tau_u \Delta\mu \quad (27)$$

$$x^{k+1} = x^k + \tau_X \Delta x \quad (28)$$

where $\Delta\mu$ and $\Delta x$ are the updates found after each NR, and $\tau_u$ and $\tau_X$ are their respective damping factors. $\tau_u$ and $\tau_X$ are chosen to ensure that the slack dual and primal variable are within the limits defined by (29) and (30).

$$\tau_u = \min\left(1, -\gamma_\mu \frac{\mu^k}{\Delta\mu}\right) \quad (29)$$

$$\tau_X = \min\left(1, \overline{x} - \gamma_X \frac{x^k}{\Delta x}\right) \quad (30)$$

where $\gamma_X, \gamma_\mu$ are constants (usually a value of 0.95) to restrict the variables $x$ and $\mu$ from oscillating around their limits. Each inequality constraint is limited separately, thereby improving convergence characteristics [16].

### C. Energizing Local Island

The optimization framework is used online to generate feasible crank-paths while receiving local measurements to continuously update the models. At each crank-step, SUGAR-R robustly adapts a pre-defined crank-path based on local measurements and device statuses, to find a feasible adjustment to the generator ramping. As a result, the local operators at each island can energize their respective island independently from the rest of the system following a distributed approach. Furthermore, operators can ensure in a timely manner, that the decisions made in the moment are feasible and optimal.

### D. Synchronizing Multiple Islands

After operators ensure proper re-energization of individual islands using SUGAR-R, they then are required to synchronize multiple islands. In real operations, synchronizers ensure that the complex voltage magnitude, phase and frequencies at the interconnection points are within operational bounds to minimize energy flow during synchronization [5]. In this simulation procedure, we strive to allow zero energy flow between the interconnection by matching the voltages, angles and frequencies prior to synchronization.

To ensure that the voltages and frequencies at the two islands are matched, the synchronization is often performed near a generating station where the operator can control both the voltage and frequency by adjusting the flow of reactive and active power as well as voltage set point [5]. Of course, operators would prefer not to make a large change in the existing frequency and voltage setpoints to reduce the chance of further system instability. Hence, to minimize the operator adjustment during synchronization, SUGAR-R aids in minimally adjusting the active and reactive power set points of participating generating stations ($\forall g \in syn$) to ensure proper synchronization of the islands (i.e. ensure complex voltages and frequency at interconnecting points are equal).

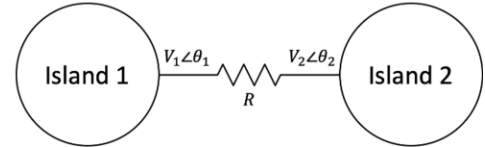

*Figure 5: Setup for simulating synchronization of two islands.*

Given two energized islands that are attempting to synchronize, we can expect one island (island 1, without a loss of generality) to match the states at the interconnection point of the other island (island 2). This entails select group of participating generators ($\forall g \in syn$) (usually one or two) in island 1 to adjust their set-points to ensure proper synchronization. SUGAR-R is able to recommend changes for proper synchronization using the approach outlined from (3)-(16) along with some additional constraints and modification to the objective function. Additional equality constraints are required to match the states of islands 1 and 2 at the point of intersection as outlined by (31)-(33). These equations are part of the dispatch for island 1 and considers the states from island 2 ($V_{R2}, V_{I2}, \Delta f_2$) are constants. This enables a distributed synchronization approach with island 2 sending minimal information to island 1 (voltage, angle and frequency at the point of intersection) through a cloud or local communication network, as shown by Fig. 1. Smaller amount of information being sent during a SCADA-absent blackout is vital as many forms of commination are unavailable.

$$V_{R1} = V_{R2} \quad (31)$$

$$V_{I1} = V_{I2} \quad (32)$$

$$\Delta f_1 = \Delta f_2 \quad (33)$$

Moreover, during synchronization study, majority of the generators (except those participating in synchronization, $(\forall g \in syn)$) operate in constant voltage operation. This requires addition of newer constraints to the problem formulation ($V_i^{SQ} = V_{i,set}^{SQ}$). However, as this optimization problem may yield an infeasible solution, we allow some deviation in the generator set points if the network is infeasible ($\Delta V_i^{SQ}$).

$$V_i^{SQ} - V_{i,set}^{SQ} + \Delta V_i^{SQ} = 0 \qquad \forall i \in PV \quad (34)$$

Operators would prefer a solution without a change of PV generator set points. To reflect this, the deviation from the voltage set-point ($\Delta V_i^{SQ}$) is highly penalized in the objective function as shown below:

$$\min_{\Delta P_{g(\forall g \in syn)}, \Delta V_i^{SQ}{}_{(\forall i \in PV)}} \left\| \Delta P_g^G \right\|_2^2 + \left\| W \Delta V_i^{SQ} \right\|_2^2 \quad (35)$$

where $W$ is a large magnitude weight that incentives PV behavior ($\Delta V_i^{SQ} \to 0$) over minimization of change in power set-points of participating generators. In this study, $\Delta V_i^{SQ}$ parameter value of greater than epsilon (say 1e-4) will direct the operator to probable reactive power and voltage issues thereby suggesting a need for broader change in grid voltage profile or PV bus setpoint prior to synchronization.

## IV. SUGAR-R FRAMEWORK

Fig. 6 shows the SUGAR-R's algorithm that assists in restoration of critical collapsed grid sections following a blackout.

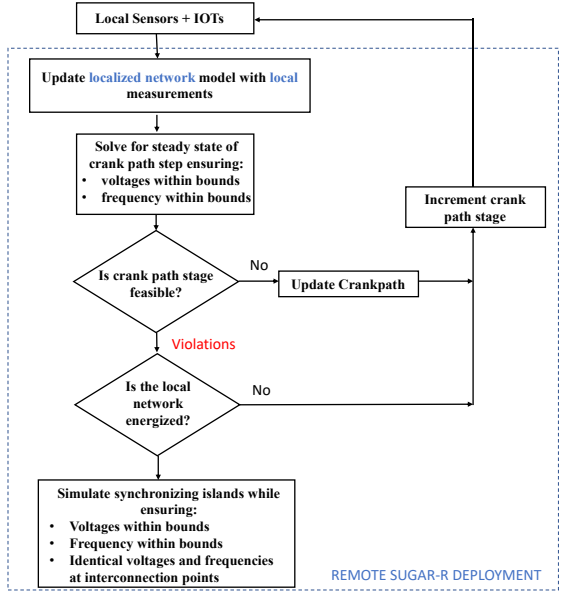

*Figure 6: Flowchart for SUGAR-R methodology for restoring grid.*

SUGAR-R framework maintains the localized network model of the grid that is being continuously updated in real-time through sampled data streams of local measurements and sensor output. Following a blackout, the framework uses the most recent network model for further analysis. The restoration analysis initiates with analyzing the first step of the crank path prior to its actual energization. If the analysis results determine that the actuation of crank path violates no device limits and results in a feasible system state with no network constraint violations, then a go-ahead is given for that crank-path to be actuated. However, if the simulation results show that no feasible grid state exists without violating device limits or tripping protective devices, then corrective actions are suggested for the updating of the crank path step based on real-power infeasibility information. These corrective actions include ramping generators real power and changing voltage setpoints. Once these changes have been realized into the crank path settings, the actual path is energized. Following which, the localized model is updated again with the measurements, and the procedure is repeated for the next crank path step. This is performed until all the steps of crank path are performed and the localized grid is restored. In case where the multiple islanded sections are energized in parallel; as a final step, the tool assists in synchronizing the islands by recommending the control setpoints such that the real and imaginary voltages at the ends of the synchronizing branch are similar with same frequency.

## V. RESULTS

To highlight the capability of the proposed SUGAR-R framework, we assist in restoration of a collapsed 39-bus network [12] following a pre-defined crank path. The crank-path for this network has 10 stages (modified from [12]) as shown in Table 1. Each subsequent stage sequentially energizes generators to bring up parts of the network, with a reference generator at bus 30.

TABLE 1: CRANKPATH FOR IEEE 39 BUS TEST CASE.

| Seq. No. | Energized Buses | Energized Gen. Id. |
|---|---|---|
| 1 | 30, 2, 25, 37 | Gen. 8 |
| 2 | 1, 39 | Gen. 1 |
| 3 | 3, 4, 5, 6, 31 | Gen. 2 |
| 4 | 25, 26, 28, 29, 38 | Gen. 9 |
| 5 | 10, 13, 14, 32 | Gen. 3 |
| 6 | 16, 17, 18, 19, 20, 34 | Gen. 5 |
| 7 | 21, 22, 35 | Gen. 6 |
| 8 | 23, 36 | Gen. 7 |
| 9 | 33 | Gen. 4 |
| 10 | 7, 8, 9, 11, 12, 15, 24, 27 | NA |

Note: Red highlighted bus number represents the generator bus.

For the purposes of this experiment, it is assumed that SUGAR-R framework has access to real-time online network model following the methodology described in Section II. We also assume access to pre-defined generator set points corresponding to the individual crank path step. Given these inputs, SUGAR-R evaluates the feasibility of each crank path step through its optimization-based framework (see Sections III and IV) and recommends changes to ramping generators set-points when necessary. After energizing the localized grid sections, SUGAR-R assists in synchronization of two islanded networks. To meet the goals, SUGAR-R uses an accurate and robust frequency-dependent optimization framework to simulate the steady-state of each crank-path, the efficacy of which will be demonstrated in comparison to other standard steady-state methodologies.

### A. Restoration through Power Flow Tool

In this experiment, we solve for each step of the crank path using a standard power flow tool without frequency dependency and document the max. and min. voltages (amongst 39 buses) and slack generator power as shown in Table 2. Here it is seen that the obtained solution is impractical for assisting in grid blackstart as the slack generator is consuming power (negative power indicated in red in Table 2), well below its active power minimum of 0 MW in many of the crank path steps. By not simulating the change in frequency due to over-generation, the slack generator is required to consume active power rather than inject. Therefore, this approach is unlikely to assist in robust energization of the grid as it is not able to mimic the reality. Moreover, this platform does not provide awareness about network feasibility or the frequency state and hence is unlikely to prevent actuation of protective devices due to out of bound state variables.

TABLE 2: SIMULATED RESULTS FOR EACH CRANKPATH STEP IN STANDARD POWER FLOW TOOL.

| Crank-path step | $V_{max}$ [pu] | $V_{min}$ [pu] | Slack set active power $P$ [pu] |
|---|---|---|---|
| 1 | 1.06 | 1.028 | -3.07 |
| 2 | 1.064 | 1.028 | -2.027 |
| 3 | 1.051 | 0.982 | 0.849 |
| 4 | 1.06 | 0.982 | 2.5 |
| 5 | 1.072 | 0.982 | -7.31 |
| 6 | 1.075 | 0.982 | 2.01 |
| 7 | 1.076 | 0.982 | -1.78 |
| 8 | 1.074 | 0.978 | -4.5 |
| 9 | 1.067 | 0.982 | -10.65 |
| 10 | 1.076 | 0.982 | -0.182 |

### B. Restoration through Governor Power Flow SUGAR

Governor power flow platform (including some industry tools) can capture frequency deviation and therefore can avoid unrealistic system state due to slack bus model by modeling implicit primary frequency control within the framework [19]. Table 3 documents the maximum and minimum voltage values and the frequency deviation from nominal for each crank path step. Although, SUGAR governor power flow-based framework provides realistic situational awareness, in the results we still observe that the frequency state is out of allowable bounds of +/- 1.2 Hz and hence can result in actuation of under/over frequency load shedding. Specifically, in the crank path stages 1, 2, 5 and 9, the large frequency deviation is a cause for concern.

TABLE 3: SIMULATED RESULTS FOR EACH CRANKPATH STEP IN GOVERNOR BASED POWER FLOW TOOL.

| Crank-path step | $V_{max}$ [pu] | $V_{min}$ [pu] | Slack bus active power $P$ [pu] | Freq. dev. $\Delta f$ [Hz] |
|---|---|---|---|---|
| 1 | 1.065 | 1.028 | 2.5 | 3.98 |
| 2 | 1.062 | 1.028 | 2.5 | 1.40 |
| 3 | 1.061 | 0.982 | 2.5 | 0.395 |
| 4 | 1.08 | 0.982 | 2.5 | 0.623 |
| 5 | 1.072 | 0.982 | 2.5 | 1.462 |
| 6 | 1.075 | 0.982 | 2.5 | 0.0639 |
| 7 | 1.079 | 0.982 | 2.5 | 0.503 |
| 8 | 1.079 | 0.978 | 2.5 | 0.724 |
| 9 | 1.08 | 0.982 | 2.5 | 1.223 |
| 10 | 1.08 | 0.982 | 2.5 | 0.002 |

This is because of primary droop control due to which, the network settles to a steady state with a positive frequency change, indicating excess power generation in the system. This may result in likely actuation of protective device tripping during energization and therefore may result in system collapse. Improved situational awareness in this framework suggests a likely issue in the crank path; however, it does not provide feedback on how best to rectify this over frequency problem while ensuring that the other system states are bounded, and the feasibility of the state is maintained. Therefore, next we demonstrate system restoration through an optimization-based SUGAR-R framework.

### C. Restoration through Optimization-Based SUGAR-R Framework.

To ensure proper steady state values for each crank path stage while re-energizing the island, SUGAR-R introduces frequency and voltage bounds, as well as a measure of active power infeasibility, $\Delta P_G$ into the formulation. The limits on frequency and voltages used in this experiment are:

| $\Delta f_{max}$ [Hz] | $\Delta f_{min}$ [Hz] | $V_{max}$ [pu] | $V_{min}$ [pu] |
|---|---|---|---|
| 1.2 | -1.2 | 1.1 | 0.9 |

Using the optimization framework described in section II, the bounds ensure stable energization of the network as shown in at each stage of the crank path. The necessary correction to generator active power setpoints to ensure such operation is reflected by $\Delta P_G$ in Table 4.

TABLE 4: SIMULATED RESULTS FOR EACH CRANKPATH STEP IN SUGAR-R FRAMEWORK.

| Crank-path step | $V_{max}$ [pu] | $V_{min}$ [pu] | Slack P set point [pu] | Freq. dev. $\Delta f$ [Hz] | Max $\Delta P_G$ [pu] | Max $\Delta P_G$ Bus |
|---|---|---|---|---|---|---|
| 1 | 1.063 | 1.028 | 0.553 | 1.200 | -1.96 | 37 |
| 2 | 1.063 | 1.028 | 2.283 | 1.200 | -2.53 | 39 |
| 3 | 1.061 | 0.982 | 2.500 | 0.395 | 0 | - |
| 4 | 1.080 | 0.982 | 2.500 | 0.623 | 0 | - |
| 5 | 1.081 | 0.982 | 2.500 | 1.200 | -0.92 | 32 |
| 6 | 1.075 | 0.982 | 2.500 | 0.064 | 0 | - |
| 7 | 1.079 | 0.982 | 2.500 | 0.503 | 0 | - |
| 8 | 1.079 | 0.978 | 2.500 | 0.724 | 0 | - |
| 9 | 1.080 | 0.982 | 2.500 | 1.200 | -0.13 | 33 |
| 10 | 1.080 | 0.982 | 2.500 | 0.002 | 0 | - |

As seen in stages 1, 2, 5 and 7 (also seen in Fig. 7), the generator active powers are re-adjusted to control the frequency such that the ramping constraints are not violated. Similarly, the tool also recommends the optimal voltage set-points for the generators during each crank-path step actuation.

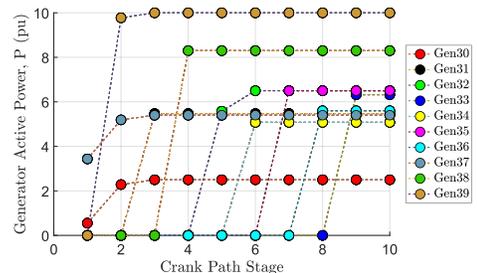

*Figure 7: Generator ramp-up during network energization.*

## D. Synchronizing Islands

In the final part of this experiment, we consider two 39-bus islanded networks that have been fully restored and require synchronization. To simulate proper synchronization at the interconnection nodes between bus 30 of island 1 (with a loading factor of 0.95) and bus 28 of island 2 (with loading factor of 1.0), we introduce three equality constraints given by (31)-(33) to result in a solution with identical complex voltages and frequency of:

$$|V_{30}^{I1}| = |V_{28}^{I2}| = 1.063 \text{ pu} \quad (36)$$

$$\theta_{30}^{I1} = \theta_{28}^{I2} = 5.34 \text{ degrees} \quad (37)$$

$$\Delta f = 0.002 \text{ Hz} \quad (38)$$

where $V_{30}^{I1}$, $\theta_{30}^{I1}$ are the voltage magnitude and angle of bus 30 on island 1, and $V_{28}^{I2}$, $\theta_{28}^{I2}$ are the voltage magnitude and angle of bus 28 of island 2. In this experiment, island 1 is adjusting the voltage and frequency at the interconnection point to match that of island 2. Two generators on island 1 (on bus 30 and bus 38 on island 1) were actively participating in aiding the synchronizing of two networks by adjusting their set points to ensure identical states on island 2 whereas other generators in island 1 operated in fixed-voltage mode. Recommended real and reactive power set-points for the participating generators to achieve similar complex voltages at the interconnection points are given below in Table 5.

TABLE 5: GENERATOR SETPOINTS FOR SYNCHRONIZATION.

| Generator | $P_G$ [MW] | $Q_G$ [MVAR] |
| --- | --- | --- |
| 30 (island 1) | 381.4 | 142.8 |
| 38 (island 1) | 373.0 | -78.4 |

## VI. CONCLUSION

In this paper, we developed a framework that assists in grid restoration under the loss of or during unavailable SCADA system. Unlike centralized grid restoration that is dependent on central SCADA and EMS, this approach works to energize critical components of the localized grid via maintaining an online network model that is continuously updated using real-time measurements. To assist in restoration, the framework simulates the online network model to ensure that each crank path step results in a feasible grid state and is actuated such that the voltage and frequency remain within bounds. To achieve this goal, the framework recommends adjustments to the generator control setpoints. After energizing individual islands, the framework assists in synchronization of islands through recommending generator setpoints that result in identical complex voltages and frequency at the interconnection points.